\documentclass[a4paper]{jpconf}
\usepackage{graphicx}
\usepackage{amsmath}
\usepackage{amssymb}

\def\e{\mathop{\rm \varepsilon}\nolimits}

\newcommand{\re}{{\rm Re}}

\newcommand{\Const}{{\rm Const.}}

\begin{document}
\title{Tunneling for a semi-classical magnetic Schr\"odinger operator with symmetries}

\author{M~Rouleux}

\address{Aix-Marseille Univ, Universit\'e de Toulon, CNRS, CPT, Marseille, France}

\ead{rouleux@univ-tln.fr}

\begin{abstract}
We are interested in decay estimates of the ground state (or the low energy eigenstates), outside the potential wells, 
for a semi-classical Magnetic Schr\"odinger operator with smooth coefficients
$P_A(x,hD_x)=(hD_x-\mu A(x))^2+V(x)$
on $L^2({\bf R}^d)$. We shall essentially consider the case
where $\mu$ is large. 
This kind of estimates, in case of
Schr\"odinger operator without a magnetic field, have been studied by Agmon \cite{Ag}, also
in the case of a Riemannian manifold $M$.
Agmon estimates hold true for any $h$,
but are particularly useful in the limit $h\to0$ when studying tunneling. 
\end{abstract}

\section{\bf Agmon estimates and tunneling parameters for the ground states}
  
Given a semi-classical Schr\"odinger operator 
$P_0(x,hD_x)=-h^2\Delta+V(x)$ on $L^2(M)$, $M$ a smooth manifold, we consider spectral effects induced by adding a 
magnetic potential (minimal coupling). For simplicity we assume here that $P_0$ and $P_A$ have compact resolvent.

A lot is known on the mathematical theory of magnetic classical and quantum Hamiltonians~: see e.g.
\cite{Iv1}, \cite{Iv2} and \cite{He2}, \cite{He3} with references therein.
The link between the classical flow 
and the spectral asymptotics, e.g. trace formulas,
is examined in \cite{Iv1}, \cite{Iv2}. This does not require precise information on eigenfunctions.
Estimates on eigenfunctions and WKB solutions considered in \cite{He2}, \cite{He3} are in turn of special importance for
investigating tunneling between ``potential wells'', 
related to the decay 
of eigenfunctions. The fact that eigenfunctions are real or complex also plays a role.
Within our framework (an application is given in \cite{IfaLouRo}),
we can summarize some of the main problems related to Magnetic Schr\"odinger operators $P_A(x,hD_x)=(hD_x-\mu A(x))^2+V(x)$
as follows.\\

\noindent $\bullet$ {\it Problem No.1}: Compare the first (non degenerate) eigenvalue $\lambda_0(h)$ of $P_0(x,hD_x)$
with the first eigenvalue $\lambda_A(h)$ (possibly degenerate) of $P_A(x,hD_x)$. 

Kato inequality shows that $\lambda_0(h)\leq\lambda_A(h)$.

More refined properties rely on the topology of $M$:

If $M$ is an open bounded subset of ${\bf R}^d$ with smooth boundary
(non necessarily simply connected), then $\lambda_0(h)=\lambda_A(h)$ iff $dA(x)=0$ in $M$ and the cohomology class
of $A$ verifies $[A]\in H^1(M,2\pi h{\bf Z})$, see \cite{LaO'Car}, \cite{He1}.
This reminds of Aharonov-Bohm effect when
$M={\bf R}^3\setminus{\bf R}(1,0,0)$ but $[A]\notin H^1(M,2\pi h{\bf Z})$,
in which case $P_A$ has continuous spectrum.

Another remarkable result (which is not to be used here) gives an upper bound on $\lambda_A(h)$. Namely, if 
$M$ is a compact manifold without boundary, then $\lambda_A(h)$ is bounded by Ma\~n\'e constant for the Lagrangian $L_A$ associated with $P_A$:
$\lambda_A(h)\leq c(L_A)$, where 
$$c(L_A)=\inf_{\phi\in C^\infty(M)}\max _{x\in M}P_A(x,{\partial\phi\over\partial x})$$
The minimizer $\phi$ is of Lipschitz class. This is investigated within ``weak KAM'' or ``weak WKB'' theory. We have also the bound
$\lambda_A(h)\leq \min c(L_A-\omega)$, over $[\omega]\in H^1(M,2\pi h{\bf Z})$, expressing in particular the gauge invariance of the spectrum, see \cite{Pa}.\\

\noindent $\bullet$ {\it Problem No.2}: Compare the decay of $u_A(h)$ with this of $u_0(h)>0$.
Generally, $\lambda_A(h)$ is simple, but $u_A(h)$ is complex and may have zeroes (vortices in 2-D, threads in 3-D, \dots). 

When $\lambda_0(h)=\lambda_A(h)$ as above, then
$u_A(x;h)=u_0(x;h)e^{i\mu(\int^x A(y)\,dy)/h}$, so the magnetic potential with $dA(x)=0$ induces no additional decay. 

Let $M$ be an open subset of ${\bf R}^d$ with smooth boundary, and $U_E=\{x\in M: V(x)\leq E\}$ be the potential well in $M$ at energy $E$.
First we assume that $U_E$ is connected. 

We focus to the case where $V(x)\geq E=0$ and $U_E=\{x\in M: V(x)=E\}=\{x_0\}$ is a non degenerate minimum (simplest case).
Actually energy level $E=0$ will be lifted by $\lambda_0(h)={\cal O}(h)$.

Decay estimates in some global $L^2$ norm with exponential weight $e^{\Phi_0(x)/h}$ for an eigenstate $u_0(h)$ of $P_0$ outside 
$U_E$ are known as {\it Agmon estimates}. They hold in any Sobolev norm and hence also pointwise. 

Let us add the vector potential $A$, with $A(x_0)=0$, and assume $\rho_0=(x_0,0)\in T^*M$ is a non degenerate elliptic point for $P_A(x,\xi)$, that
we call the {\it magnetic well}. 
 
In case of $P_A$ we know already that Agmon type estimates with the same weight $e^{\Phi_0(x)/h}$ 
hold for $u_A(h)$ \cite{HeSj2}, \cite{He1}. Thus we call them ``basic Agmon estimates''. 
But generally speaking, magnetic fields
are confining, so we can expect that the eigenfunction $u_A(h)$ will decay faster indeed than $u_0(h)$ outside $x_0$. 

Local WKB constructions of $u_A(h)$ near the magnetic well, using quadratic approximation near the elliptic point,
are computed in \cite{MaSo}. When $V,A$ are analytic, they are of the form
\begin{equation}\label{5}
  u_A(x;h)=e^{-\Psi_A(x)/h}\bigl(a(x,h)+{\cal O}(e^{-{\e }_A/h})\bigr), \ x\in B_A(x_0,r_0)
\end{equation}
and $u_A(h)$ decays generally faster than $u_0(h)$ in $B_A(x_0,r_0)$.
Using almost analytic extensions,
we can show that (\ref{5}) holds true when $A$ and $V$ are merely $C^\infty$, provided we replace
${\cal O}(e^{-{\e }_A/h})$ by ${\cal O}(h^\infty)$. 

However, it is 
difficult to extend these expansions in the larger domain $M$. 
To this end, we introduce here some {\it relative Agmon estimates}: namely, at least for sufficiently large coupling constant $\mu$,
$u_A(h)$ decays in $L^2(M)$ norm as $e^{-\Phi_1(x)/h}$ with $\Phi_1>\Phi_0$, and $\Phi_1\approx \re\Psi_A$ in $B_A$. 
\\

\noindent $\bullet$ {\it Problem No.3} (tunneling for the double well $\{x^<,x^>\}$). 
  When $M={\bf R}^d$ instead, we assume that $P_0$ and $P_A$ commute with some hyperplane symmetry $\sigma(x',x_d)=(x',-x_d)$. Let
  $M^>$ be a sufficiently large open set containing
$x^>$, but not $x^<\in M^<=\sigma(M^>)$.
  
In the presence of magnetic wells, we are interested in an
estimate (from above) of the splitting between the two first eigenvalues $\lambda^\pm_A(h)$, which, incidentally, may coincide. 

Much is known for the splitting (from above and from below) between $\lambda_0^+(h)>\lambda_0^-(h)$, see \cite{HeSj1}, \cite{Ma1}.

In case $V$ has non-degenerate minima at $x^{>,<}\in\{\pm x_d>0\}$, $V(x^{>,<})=E=0$ we can construct (quite accurately from the point of vue of
tunneling)
approximate eigenfunctions $u_0^\pm(h)$ for $P_0$ in $L^2({\bf R}^d)$ from the ground state $u_0(h)$ of the one-well problem,
i.e. localized at $x_0=0$.
Here $u_0^-(h)$ is even in $x$ and given roughly by 
$u_0^-(x;h)={1\over\sqrt2}\bigl(u_0(x-x^>;h)+u_0(x-x^<;h)\bigr)$,
while $u_0^+(h)$ is odd in $x$ and of the form
$u_0^+(x;h)={1\over\sqrt2}\bigl(u_0(x-x^>+;h)-u_0(x-x^<;h)\bigr)$
Here we have assumed for simplicity that $u_0$ is even in $x$, see \cite{Ha} for the correct formula.

Define as a {\it libration} $\gamma(h)$ a minimal geodesic for Agmon distance at energy $\lambda_0(h)$, with $\lambda_0(h)={\cal O}(h)$,  
connecting $U^{>,<}(h)=\{x\in M^{>,<}: V(x)\leq \lambda_0(h)\}$, see e.g. \cite{AnRo},
\cite{AnDoNo}. An instanton $\gamma$ is a limiting curve of the family of librations,
as $h\to0$, i.e. as $U^{>,<}(h)$ shrinks to $\{x^{>,<}\}$, the well at energy 0. This is a  minimal geodesic for Agmon distance between $x^<$ and $x^>$
at energy 0. 
Instantons form generically a discrete set.

Localizing Agmon estimates near $\gamma$ gives \cite{HeSj1}, \cite{Ma1}: for all $\delta >0$
\begin{equation}\label{14}
  \Delta\lambda_0(h)=\lambda^+_0(h)-\lambda^-_0(h)={\cal O}_{\delta} \bigl(\exp\bigl[-(d_0(x^>,x^<)-\delta )/h\bigr]\bigr)
  \end{equation}
when $0<h<h_{\delta }$. Here $d_0(x^>,x^<)$ is Agmon distance between $x^>$ and $x^<$, that satisfies eikonal equality
$|\nabla d_0(x^<,x)|^2=|\nabla d_0(x,x^>)|^2=V(x)$ (here we use the stable/unstable manifold theorem).
We shall content ourselves with leading exponential estimates, allowing for arbitrary small loss
$\delta $ in the exponent. 

Introduce now the magnetic potential. Again the localization procedure works, and it suffices to know the
ground state $u_A(h)$ of the one-well problem. 

For very small ($h$ dependent) coupling constant $\mu$, WKB expansions of $u_A(h)$
yield an estimate similar to (\ref{14}) on the splitting $\Delta\lambda_A(h)$ \cite{HeSj2}. For small coupling constant $\mu$ (but independent of $h$)
we need analyticity of $V,A$ 
to ensure exponential accuracy on WKB solutions. 

Under some suitable hypotheses,  
we expect that  our relative Agmon estimates approximate $\Delta\lambda_A(h)$ in a similar way to (\ref{14}), and moreover
  \begin{equation}\label{15}
    \liminf_{h\to0}\bigl(-h \log \Delta\lambda_A(h) +h \log \Delta\lambda_0(h)\bigr)>0
    \end{equation}
  This holds actually in 2 regimes: (1) the coupling constant $\mu$ is sufficiently large, so that the decay of $u_A(h)$ is little sensitive to $V$;
  (2) $\mu^2|A(x)|^2-V(x)$ is positive and small enough.
  
Recent results in the 2-D case of compactly supported and radially symmetric potential wells, and constant $B$
(\cite{FeShaWe}, \cite{HeKa}), express the rate of exponential decay in term
of the {\it hopping formula}. See also \cite{HeKaSu}, and \cite{FoMoRa} when $V=0$. 

\section{\bf Precise statements and outline of proofs}

\noindent {\it 1) General Agmon identities in the case of a magnetic potential}\\
  
The scalar and magnetic potentials $V,A$ will be assumed to be $C^\infty$.
Let $\Omega\subset{\bf R}^d$ be a bounded open set with $C^2$ boundary.

We will consider weighted integrals, related with Dirichlet forms, like
$\int_\Omega|(hD_x-\mu A(x)) e^{\Phi(x)/h}u|^2\,dx$ or
$\int_\Omega|(hD_x-\mu A(x)+ihf(x)) e^{\Phi(x)/h}u|^2\,dx$
for a suitable real vector field $f$. 

When $u$ is a complex $C^2$ function on $\Omega$, we define the real vector field (quantum current with $U(1)$ symmetry)
$$J_u=-i(\nabla u\cdot\overline u-\nabla\overline u\cdot u)$$
It is not a priori
integrable on $\Omega$ when $\Omega$ is not simply connected but if $u$ is not vanishing, then
$J_u=|u|^2\nabla\arg u^2$.
Using Green formula one can show:\\

\noindent {\bf Proposition 1}: {\it Let $\Phi=\Phi_R+i\Phi_I$ be a complex Lipschitz function defined on $\Omega$. 
For all $u\in C^2(\overline\Omega;{\bf C})$, we have 
\begin{equation}\label{1}
  \begin{aligned}
&\int_\Omega|(hD_x-\mu A(x)) e^{\Phi(x)/h}u|^2\,dx+\\
&+\int_\Omega e^{2\Phi_R(x)/h}\bigl(V(x)-|\nabla\Phi_R(x)|^2-|\nabla\Phi_I(x)|^2+2\mu\langle\nabla\Phi_I(x),A(x)\rangle\bigr)|u(x)|^2\,dx\\
&-h\int_\Omega e^{2\Phi_R(x)/h}\langle J_u(x),\nabla\Phi_I(x)\rangle\,dx=\\
&\re\int_\Omega e^{2\Phi_R(x)/h}P_A(x,hD_x)u(x)\cdot \overline u(x)\,dx+h^2\re\int_{\partial\Omega} e^{2\Phi_R/h}\overline u{\partial u\over\partial\nu}
    \,d\sigma
    \end{aligned}
\end{equation}
where $\nabla\Phi\in L^\infty(\Omega)$
is understood in the usual sense, and $d\sigma$ is the surface measure on $\partial\Omega$.}

As a particular case $\Phi_I=0$, we retrieve the classical formula \cite{He1}, p.95 which will give the ``basic'' Agmon estimate~:\\

\noindent {\bf Corollary 2}: {\it Let $\Phi=\Phi_R$ be a Lipschitz real function defined on $\Omega$. 
For all $u\in C^2(\overline\Omega;{\bf C})$, we have }
\begin{equation}\label{2}
  \begin{aligned}
\int_\Omega&|(hD_x-\mu A(x)) e^{\Phi(x)/h}u|^2\,dx
+\int_\Omega e^{2\Phi(x)/h}\bigl(V(x)-|\nabla\Phi(x)|^2\bigr)|u(x)|^2\,dx\\
&=\re\int_\Omega e^{2\Phi(x)/h}P_A(x,hD_x)u(x)\cdot \overline u(x)\,dx+
h^2\re\int_{\partial\Omega} e^{2\Phi(x)/h}\overline u{\partial u\over\partial\nu}(x)
\,d\sigma(x)
  \end{aligned}
  \end{equation}

\noindent {\it 2) Standard Agmon estimate without a magnetic potential: a review}\\

Assume also $A=0$, and consider Dirichlet realization of $P_0(x,hD_x)=-h^2\Delta+V(x)$ in a domain $M\subset{\bf R}^d$.
For simplicity we assume $M$ to be bounded.
Let $u_0(h)$ be a normalized eigenfunction,
associated with the eigenvalue $\lambda_0(h)\in ]0,Ch]$. We apply Corollary 2 to $\Omega=M$. 
Assume here that $V$ has a single potential well
$U_0(h)=\{x\in M: V(x)\leq \lambda_0(h)\}$ at energy $\lambda_0(h)$, let $\Phi_0(x)=d_0(x,U_0(h))$
be Agmon distance of $x$ to $U_0(h)$, i.e. satisfy the differential inequality (only when ``$\lambda_0(h)=0$'' we have an equality)
\begin{equation}\label{11}
  |\nabla\Phi_0(x)|^2\leq \bigl(V(x)-\lambda_0(h)\bigr)_+
  \end{equation}

Then Corollary 2 for $A=0$  implies the following standard Agmon estimate. Namely, there is $a(\delta)>0$, with $a(\delta)\to0$ as $\delta\to 0$, such that
\begin{equation}\label{3}
  h^2\int_M|\nabla e^{\Phi_0(x)/h}u_0(x;h)|^2\,dx+\int_M e^{2\Phi_0(x)/h}|u_0(x;h)|^2\,dx\leq C_\delta e^{2a(\delta)/h}
  \end{equation}
for $0<h\leq h_\delta$. Using the equation $P_0(x,hD_x)u_0(h)=\lambda_0(h)$, this $L^2$ estimate carries to any Sobolev space
$H^s(M)$, and by Sobolev embedding theorem, also to $L^\infty$ estimates.

For the symmetric double well problem in a ``large'' bounded domain $M\subset{\bf R}^d$,
with a single well $U^{>,<}(h)$ on each side of the hyperplane $x_d=0$.
consider instead Dirichlet realization of $P_0(x,hD_x)=-h^2\Delta+V(x)$ ,
containing sub-domains $M^{>,<}\subset M$ as above
symmetric of each other with respect to $x_d=0$, $U^{>,<}(h)\subset M^{>,<}$, but $U^>(h)\cap M^<=\emptyset$.

We set $\widetilde\Phi_0(x)=\min\bigl(d_0(x,U^>(h)),d_0(x,U^<(h))\bigr)$. 

Then (\ref{3})  holds true for $\widetilde\Phi_0(x)$ instead of $\Phi_0(x)$, see also \cite{HeSj1}, \cite{He1} when $M={\bf R}^d$.\\

\noindent {\it 3) Standard Agmon estimates with a magnetic potential}

Consider next the Dirichlet realization of $P_A(x,hD_x)=(hD_x-\mu A(x))^2+V(x)$ in the domain $M\subset{\bf R}^d$, $\lambda_A(h)$ 
its ground state energy (possibly degenerate), and 
$u_A(h)$ an eigenfunction associated with $\lambda_A(h)$. 
As before we start with the ``magnetic one-well problem''. 
Let $\Omega$ be the complement of a (euclidian) ball $B_A\subset M$ (the projection of the ``magnetic well'').

We have the ``basic'' Agmon estimate for $u_A(x;h)$, which we express as in (\ref{3}). 
Namely for all $\delta>0$, there is $C_\delta>0$ such that, uniformly for $0<h\leq h_\delta$
\begin{equation}\label{4}
  \begin{aligned}
&\int_\Omega|(hD_x-\mu A(x)) e^{\Phi_0(x)/h}u_A(x;h)|^2+\int_\Omega e^{2\Phi_0(x)/h}|u_A(x;h)|^2\leq\\
&C_\delta \sup_{\Omega}e^{2\delta\Phi_0/h}\int_{\partial\Omega} e^{2(1-\delta)\Phi_0(x)/h}|\overline u_A{\partial u_A\over\partial\nu}|
  \end{aligned}
  \end{equation}
  
At this point we use the precise local decay estimates (\ref{5}) of $u_A(h)$
to control the RHS of (\ref{4}). 
Recall 
$\Psi_A$ is a (possibly complex) phase with real part that we shall assume strictly larger than $\Phi_0$ on $\partial B_A$, at least for large $\mu$.
This assumption is supported by the following example on $L^2({\bf R}^2)$ 
\begin{equation}\label{9}
  P_A(x,hD_x)=(hD_{x_1}+{\mu\over2}x_2)^2+(hD_{x_2}-{\mu\over2}x_1)^2+x_1^2+x_2^2
  \end{equation}
with ground state of the form $u_A(x;h)=\Const \exp[-c(\mu)x^2/2h]$, and ground state energy
$\lambda_A(h)=c(\mu)h=(1+\mu^2/4)^{1/2}h$. Actually this extends easily to higher dimensions, see \cite{Sh}.  

After some manipulation, (\ref{4}) reduces to an estimate like (\ref{3}) where the RHS can be replaced by ${\cal O}(e^{-\e _1h})$,
for some $\e _1>0$, 
provided $u_A(x;h)$ decreases indeed faster (as in the latter example) than $e^{-\Phi_0(x)/h}$ on $\partial B_A$. \\

\noindent{\it 4) R.Lavine and M.O.Carroll's formula and relative Agmon estimate}

Contrary to (\ref{2}) that suggests to introduce Agmon distance $d_0$ as a natural geometric guideline, (\ref{1})
encodes no direct geometric information for complex $\Phi$. This reflects the intrication between
the cyclotronic motion and the magnetic drift at the classical level. We only know \cite{Iv2} that these motions are
relatively well decoupled for large $\mu$. Thus we can expect that in this regime to factor out the wave function
as $u_A(h)=u_0(h)v(h)$ and find Agmon type estimates on $v(h)$.
Recall the following formula \cite{LaO'Car}: 

Let $A,f$ be real vector fields locally square integrable on ${\bf R}^d$, then for all
$\widetilde u\in C^\infty_0({\bf R}^d;{\bf C})$ we have 
\begin{equation}\label{6}
  \int_{{\bf R}^d}|(hD_x-\mu A(x)+ihf) \widetilde u|^2\,dx=\int_{{\bf R}^d}|(hD_x-\mu A(x)) \widetilde u|^2\,dx+h^2\int_{{\bf R}^d}
  (\nabla f+|f|^2)|\widetilde u|^2\,dx
  \end{equation}
\smallskip
Let $(\lambda_0(h), u_0(h))$ be the ground state of the Dirichlet realization of $P_0$ in $M$, then 
$f=\nabla u_0/u_0$ is a natural choice, for
\begin{equation*}
  h^2\nabla\cdot{\nabla u_0\over u_0}=h^2{\Delta u_0\over u_0}-h^2\big|{\nabla u_0\over u_0}\big|^2=V-\lambda_0(h)-h^2\big|{\nabla u_0\over u_0}\big|^2
  \end{equation*}
so that $h^2(\nabla f+|f|^2)|u|^2=(V(x)-\lambda_0(h))|u|^2$.

Let as above $\lambda_A(h)$ (we assume to be non degenerate) be the ground state energy of
of $P_A(x,hD_x)$ on $L^2(M)$ with Dirichlet boundary condition, and $u_A(h)$ the corresponding normalized eigenfunction, which we write as $u_A(h)=u_0(h)v(h)$.
Suitably modifying ({\ref 6}) with $\widetilde u=e^{\Phi/h}u_A(h)$ to account for boundary terms on $\Omega\subset M$,
we can express $\int_{{\bf R}^d}|(hD_x-\mu A(x)+ihf) \widetilde u|^2\,dx$ in term of a Dirichlet form
$\int_\Omega e^{2\Phi_R(x)/h}u_0^2(x;h)|(h\nabla+\nabla\Phi_R)v(x;h)|^2\,dx$ and get~:\\

\noindent {\bf Proposition 3}: {\it Let $\Phi_R,\Phi_I$ be Lipschitz real functions defined on $\Omega$. 
If $u_A(h)=u_0(h)v(h)$, we have }
\begin{equation}\label{7}
\begin{aligned}
&\int_\Omega e^{2\Phi_R(x)/h}u_0^2(x;h)|(h\nabla+\nabla\Phi_R)v(x;h)|^2\,dx+\\
&\int_\Omega e^{2\Phi_R(x)/h}u_0^2(x;h)\bigl(-|\nabla\Phi_R(x)|^2+|A(x)|^2-\lambda_A(h)+\lambda_0(h))\bigr)|v(x;h)|^2\,dx-\\
&h\int_\Omega e^{2\Phi_R(x)/h}u_0^2(x;h)\langle J_v(x),2\nabla\Phi_I(x)-A(x)\rangle\,dx=\\
&h^2\re\int_{\partial\Omega} e^{2\Phi(x)/h}\overline u_A{\partial u_A\over\partial\nu}(x)
-h^2\int_{\partial\Omega} e^{2\Phi_R(x)/h}|u_A|^2\langle f,\nu\rangle
\end{aligned}
\end{equation}

To get rid of the extra term on the RHS we try to solve
\begin{equation}\label{8}
  \langle J_v(x),2\nabla\Phi_I(x)-A(x)\rangle=0, \quad x\in M
  \end{equation}
by
$$2\langle J_v(x),\nabla\Phi_I(x)\rangle=\langle J_v(x),A(x)\rangle$$
This can be done if $J_v(x)\neq0$ is an integrable field in the simply connected open set $M$.
When $u_A(h)$ is non vanishing in $M$, $J_v(x)=|u_A(x;h)|^2\nabla\arg u_A^2(x;h)$,
and $\arg u_A^2(x;h)$ is a smooth function. 
Vanishing of $u_A$ maybe an obstruction for solving (\ref{8}), in particular when the zeroes $x_j$ of $J_v(x)$ are vortices. 
We can also try to solve (\ref{8}) approximately along a magnetic line $\gamma$, namely look for
$\Phi_I$ such that the 1-form
$2d\Phi_I(x)-\mu A(x)$ vanishes on $\gamma$. 

Once the contribution of $J_v(x)$ in (\ref{8}) has been removed, we are left with an expression similar to (\ref{2}),
with $|A(x)|^2$ instead of $V(x)$. Provided
$A\neq0$ outside the well, we solve an eikonal inequality
\begin{equation}\label{12}
  |\nabla\Phi_1|^2\leq\bigl(\mu^2|A(x)|^2-\lambda_A(h)+\lambda_0(h)\bigr)_+
  \end{equation}
and put $\Phi_R=(1-\delta)\Phi_1$. This yields again an Agmon estimate of the form (\ref{4}), with $\Phi_1$ instead of $\Phi_0$. 
On the other hand, we know that $u_0(x;h)$ is everywhere of the same order of magnitude as $e^{-\Phi_0(x)/h}$.
More precisely, 
if $0\leq V\in C^\infty(M)$ and $u_0$ be the normalized ground state of $P_0=-h^2\Delta+V(x)$ on $L^2(M)$, then for all $\delta >0$,
$u_0(x;h)\geq C_{\delta }e^{-(d_0(x,U_h)+\delta )/h}$
locally uniformly on any compact set $K\subset M\setminus U_0(h)$, see e.g. \cite{MaRo}; moreover
in a neighborhood of the quadratic well $x_0$, we have
$|u_0(x;h)|\leq C h^{-N}e^{-\Phi_0(x)/h}$
(as well as all derivatives), see \cite{He1}, Proposition 3.3.5. 

In turn, this gives again an estimate like (\ref{3}), so there is a set of two (independent) Agmon estimates for $u_A(h)$.
Let $\Phi_A(x)=\max (\Phi_0(x),\Phi_1(x))$, which amounts to consider the
(Lipschitz) distance $d_A(x)=\max (d_0(x,U_0(h)),d_1(x,U_1(h))$, where
$U_0(h)=\{x\in M: V(x)-\lambda_0(h)\leq0\}$,
$U_1(h)=\{x\in M: \mu^2|A(x)|^2-\lambda_A(h)+\lambda_0(h)\leq0\}$. 
Thus we proved:\\

\noindent {\bf Theorem 4}: {\it Let $u_A(h)$ be an eigenfunction of $P_A$ on $L^2(M)$  
with eigenvalue $\lambda_A(h)$. Assume we have solved (\ref{8}). Then 
\begin{equation}\label{10}
  \int_\Omega|\nabla e^{\Phi_A(x)/h}u_A(x;h)|^2\,dx+\int_\Omega e^{2\Phi_A(x)/h}|u_A(x;h)|^2\,dx\leq C_\delta e^{2a(\delta)/h}
\end{equation}
with $a(\delta)\to0$ as $\delta\to0$. Moreover, if the real part of $\Psi_A(x)=\Psi_A(x,\mu)$ is increasing with $\mu$
on $\partial B_A(x_0,r_0)$ as in Example (\ref{9}), then
the remainder term ${\cal O}(e^{2a(\delta)/h})$ can be improved to ${\cal O}(e^{-\e _2/h})$ for $\mu$ large enough, $\e _2>0$. }\\

This is consistent with (\ref{9}) where $\Psi_A=\sqrt{\Phi_0^2+\Phi_1^2}\geq\Phi_A$. But it is clear that our method cannot capture the precise decay
of (\ref{9}). 

The point is that in general $d_A$ is no longer a (degenerate) Riemannian metric.  
This situation is met however in the following cases, discarding the energy shifts 
$\lambda_0(h)$ or $\lambda_A(h)-\lambda_0(h)$ in the eikonal inequalities (\ref{11}) and (\ref{12})~: (1)
$\Phi_A=\Phi_1$ almost everywhere (i.e. the magnetic potential is more confining
than the scalar potential); (2) the potentials verify $\mu^2|A(x)|^2=V(x)$, or $\mu^2|A(x)|^2-V(x)$ is sufficiently small. The latter
case can be treated by perturbation theory, since the trajectories for $\xi^2-V(x)$ at energy $\lambda_0(h)$  and $\xi^2-|A(x)|^2$
at energy $\lambda_A(h)-\lambda_0(h)$ will be comparable, in particular near the set of minimal geodesics of either system.
\\

\section{\bf Application to tunneling}

We present here the main ideas towards (\ref{15}). As usual \cite{HeSj2} we need first an assumption on the spectrum (spectral gap).
Let $\lambda_A^\pm(h)$ be the two first eigenvalues of $P_A$ on $L^2({\bf R}^d)$, $\mu_A(h)=\mu_A^>(h)=\mu_A^<(h)$  this of the Dirichlet realization
of $P_A^{M^\pm}$ on the localized domains $M=M^{>,<}$ as above.  Assume $\mu_A(h)$
is simple and asymptotically simple, see e.g. \cite{MaRo}. As a rule, we expect
the splitting $\Delta\lambda_A(h)=\lambda_A^+(h)-\lambda_A^-(h)$
to be ${\cal O}(e^{-S_A/h})$ for some ``Agmon distance'' $S_A>0$ between the magnetic wells, to be computed
in term of the ground state $u^{>,<}$ of $P_A^{M^{>,<}}$, see \cite{HeSj2},
Theorem 3.1 and Remark 3.7. The ``gap formula'' is thus of the form
\begin{equation}\label{13}
  \begin{aligned}
  \Delta&\lambda_A(h)=h^2\int_\Gamma \bigl(\, \overline {u^>}(y)\partial_n u^<(y)-u^<(y)\partial_n \overline{u^>}(y)\bigr)\,dS(y)+\\
  &h\int_\Gamma \bigl(\, \overline {u^>}(y)\langle A(y),n(y)\rangle u^<(y)
  -\overline{u^<}(y)\langle A(y),n(y)\rangle u^>(y)\bigr)\,dS(y)
  \end{aligned}
\end{equation}
where $\Gamma\subset\{x_d=0\}$ is a ``geodesic bisector'' and
$n$ a normal to $\Gamma$. The eigenvalues $\lambda_A^\pm(h)$ are exponentially close to the corresponding
$\mu_A(h)$, and the functions $u^>,u^<$ are related by the hyperplane symmetry.\\

In the case $\Phi_A=\Phi_1$, we apply Theorem 4 to the functions $u^>, u^<$, which allows to bound from above 
the first two traces of $u^>,u^<$ in $L^2(\Gamma)$
with the exponential weight $e^{-\Phi_A/h}$. This is essentially done as in the situation without the magnetic potential. 
This gives  $\Delta\lambda(h)={\cal O}(e^{-(S_A-c)/h})$, $c$ arbitrary small, where $S_A$ is
(ordinary) Agmon distance between the wells $x^>$ and $x^<$ for the potential $|A(x)|^2$.

The case $\e W(x)=\mu^2|A(x)|^2-V(x)$ small (where we can set $\mu=1$) is more difficult since we must work directly from (\ref{7}).
We can replace $\Omega$ by a neighborhood $\Omega_1$ of a minimal geodesic between
$x^>$ and $x^<$ common to $|A(x)|^2$ and $V(x)$
when they are equal, and use perturbation theory in $\e $. We expect again
$\Delta\lambda_A(h)={\cal O}(e^{-(S_A-c )/h})$, $c$ arbitrary small,
where $S_A$ is, at 0-order in $\e $, Agmon distance between the wells $x^{>}$ and $x^<$ for the potential $|A(x)|^2=V(x)$.
Moreover when $W(x)>0$, $\Phi_R$ should be strictly larger than $\Phi_0$ in $\Omega_1$, which implies (\ref{15}).\\

\noindent {\it Acknowledgements}: I thank Bernard Helffer for his useful remarks and interesting references. \\

\end{document}